\documentstyle[12pt]{article}
\begin{document}
\vskip 2cm
\begin{center}
{\Large  {\bf A novel observation in the BRST approach to a free spinning relativistic particle }}

\vskip 3.0cm

{\sf S. Krishna$^{(a)}$, D. Shukla$^{(a)}$, R. P. Malik$^{(a,b)}$}\\
$^{(a)}$ {\it Physics Department, Centre of Advanced Studies,}\\
{\it Banaras Hindu University, Varanasi - 221 005, (U.P.), India}\\

\vskip 0.1cm

{\bf and}\\

\vskip 0.1cm

$^{(b)}$ {\it DST Centre for Interdisciplinary Mathematical Sciences,}\\
{\it Faculty of Science, Banaras Hindu University, Varanasi - 221 005, India}\\
{\small {\sf {e-mails: skrishna.bhu@gmail.com; dheerajkumarshukla@gmail.com;   malik@bhu.ac.in}}}

\end{center}

\vskip 2cm

\noindent
{\bf Abstract:} 
For the newly proposed coupled (but equivalent) Lagrangians for the supersymmetric (SUSY) system of a one 
(0 + 1)-dimensional spinning relativistic particle, we derive the Noether conserved charges corresponding 
to its (super)gauge, Becchi-Rouet-Stora-Tyutin (BRST), anti-BRST and ghost-scale symmetry transformations. 
We deduce the underlying algebra amongst the continuous symmetry operators and corresponding conserved 
charges. We point out a novel observation that emerges, for this specific SUSY system, when we discuss it 
within the framework of BRST formalism. In particular, the requirement of the physicality criteria with the (anti-)BRST charges 
leads to a completely new observation because, as it turns out, one of the primary constraints does not annihilate the physical state 
of the theory. We have never come across this kind of result in the realm of BRST approach to 
{\it usual} gauge theories. 
We provide the physical and theoretical reasons behind the above observation.


\newpage

\newpage

\section{Introduction}

The model of a   spinning relativistic particle (SRP) presents a prototype 
example of a one (0 + 1)-dimensi-onal (1D) supersymmetric (SUSY) system (embedded  in a  $D$-dimensional 
Minkowskian target spacetime supermanifold) which is characterized by the bosonic variables 
$x^\mu (\tau) (\mu = 0, 1, 2,.....D-1)$ and its fermionic (i.e. SUSY) partners
$\psi^\mu (\tau)  (\mu = 0, 1, 2, ....D-1 )$ where $\tau$ is a monotonically increasing evolution 
parameter that describes  the super world-line traced out by the motion of the SPR. 
This model is interesting as it respects the (super)gauge symmetries as well as the reparametrization 
symmetry in a very elegant manner. As a consequence, it provides a toy model for the SUSY gauge theories as well as
supergravity theories. Its generalization leads to the emergence of reparametrization 
invariant models for the (super)strings, too (see, e.g. [1, 2] for details).

The model of SRP has been studied from many angles (see, e.g. [3-6]). The BRST analysis of this model has also
been performed in [6] by exploiting its (super)gauge symmetries. However, only its BRST symmetries 
have been discussed and its proper anti-BRST transformations have been left untouched. In a 
very recent paper [7], we have exploited the theoretical arsenal  of the augmented version of 
superfield formalism [8-11] to obtain a consistent  set of (anti-)BRST transformations. 
The existence of anti-BRST symmetry is important as it has its mathematical origin in the concept of gerbes [12,13].
The full set of proper (i.e. off-shell nilpotent and absolutely  anticommuting) (anti-)BRST symmetries 
are shown to be respected  by a coupled (but equivalent) set of Lagrangians 
for our present theory (see, e.g. [7] for details).

The purpose  of our present investigation is to discuss various continuous symmetries of the above coupled
Lagrangians and derive corresponding conserved charges. We also discuss the underlying algebra amongst the 
symmetry operators and their corresponding charges in an explicit fashion. We demonstrate some novel 
features in the context of physicality condition for this model which have not been observed, hitherto, 
in the context of application of BRST formalism to any other gauge/reparametrization invariant theories.
It turns out that the requirements of the annihilation of the physical states by the conserved (anti-)BRST charges
do not produce the annihilation of these physical states by one of the primary constraints of the theory.
This is a completely new observation in the realm  of application of BRST formalism to a physical system
with constraints.

One of the other novel observations for this model is the existence of Curci-Ferrari (CF)-type restriction 
which is the root cause behind the existence  of coupled (but equivalent) Lagrangians as well as the absolute 
anticommutativity of the (anti-)BRST symmetries (and  corresponding conserved charges). This restriction
has been derived  by exploiting the basic tenets of superfield approach [8-11] to BRST formalism.
However, this restriction also emerges as an off-shoot   of  the equations of motion from  the coupled 
(but equivalent) Lagrangians for our present system (cf. (8) below).

The central motivating factors behind our present investigation are as follows. First and foremost, we observe that the physicality 
criteria with conserved and nilpotent (anti-) BRST charges do not select out the physical state that is annihilated by one 
of the primary constrains of the theory. Second, within the framework of BRST formalism, for the first-time, we observe that the above specific
primary constraint is expressed in terms of the (anti-)ghost variables of the theory          
which is totally different from our experience in usual gauge theories. Third, our present attempt would, perhaps, provide some insights 
into application of BRST formalism to supergravity theories. Finally, our present investigation is our modest 
first-step towards our main goal of applying the BRST formalism to supersymmetric $p$-form ($p = 1, 2, 3,...$) (non-)Abelian gauge theories.

Our present paper is organized as follows. In our Sec. 2, we recapitulate  the bare
essentials of (super) gauge symmetries and their generalizations to the (anti-)BRST symmetry 
transformations in the Lagrangian formulation where a coupled (but equivalent) set of the latter exist. 
Our Sec. 3 is devoted to a brief discussion of the constraints of our present SUSY system of 
SRP and generators constructed from them. In our Sec. 4, we dwell
on the derivation of BRST charge and establish  its superiority over the generator of our previous 
section. Our Sec. 5 deals with the derivation of anti-BRST charge. The ghost charge and 
the complete  set of BRST algebra are deduced in our Sec. 6. Finally, we make some concluding 
remarks in our Sec. 7  where we compare and contrast the importance of (anti-)BRST charges
and the generators (constructed from the first-class constraints of our present theory).

\section{Preliminaries: (Super)gauge  and (anti-)BRST symmetries in the Lagrangian formalism}

\label{sec:1}
We begin with the following first-order Lagrangian ($L_f$) for a one (0 + 1)-dimensional supersymmetric
system of a massive spinning relativistic particle [6] 
\begin{eqnarray}
L_f = p_\mu\; \dot x^\mu - \frac{e}{2}\; (p^2- m^2) + \frac{i}{2}\; (\psi_\mu \;\dot \psi^\mu - 
\psi_5 \;{\dot\psi}_5) + i \;\chi\; (p_\mu\; \psi^\mu - m\;\psi_5),
\end{eqnarray}
where ${\dot x}^\mu = (dx^\mu/d\tau)$ and ${\dot \psi}^\mu = (d\psi^\mu/d\tau)$ are the generalized 
``velocities'' for the target space bosonic variable $x^\mu$ and its fermionic ($\psi_\mu \psi_\nu 
+ \psi_\nu \psi_\mu = 0,\;\; \psi^2_\mu = 0$) counterpart $\psi^\mu$ where $\tau$
is the parameter that describes the  evolution of our present SUSY system. The canonical momenta 
$\Pi_e = ({\partial L_f}/{\partial\dot e}) \approx 0$ and $\Pi_\chi = ({\partial L_f}/{\partial\dot \chi})
\approx 0$ are the primary constraints on the theory  and the secondary constraints $(p^2 - m^2) \approx 0$ 
and $(p_\mu \psi^\mu -  m \psi_5 ) \approx 0$ emerge out from the equations of motion corresponding to the 
Lagrange multiplier bosonic and fermionic variables $e(\tau)$ and $\chi(\tau)$, respectively. These 
variables are the analogues of the vierbein and Rarita-Schwinger fields of the 4D supergravity theories. 
For our present SUSY theory, the above multiplier variables are the gauge- and super-gauge variables, too.
All the fermionic variables (e.g. $\psi_\mu, \chi$)  anticommute ($\psi_\mu \psi_\nu + \psi_\nu \psi_\mu = 0,
\; \psi_\mu \chi + \chi \psi_\mu = 0,\; \psi^2_\mu = 0, \;\chi^2 = 0$) among themselves and they commute 
(i.e. $\psi_\mu x_\nu - x_\nu \psi_\mu = 0,\;  \psi_\mu e - e \psi_\mu = 0,\; \chi x_\mu - x_\mu \chi = 0, $ etc.)
with all the bosonic variables of our present SUSY theory. In addition to the above variables, a fermionic
($\psi_5 \;\psi_\mu + \psi_\mu\; \psi_5 = 0, \;\psi_5 \;\chi + \chi \;\psi_5 = 0,$ etc.) variable $\psi_5(\tau)$
has been introduced to take care of the mass $m$ of the SUSY particle. The variable $\psi_5 (\tau)$ satisfies: 
$\{\psi_5, \psi_5 \} = 1$ as well as $\psi_5 \;x_\mu - x_\mu \;\psi_5 = 0,\; \psi_5\; e - e\; \psi_5 = 0$, etc.

The above Lagrangian is endowed with the following local, continuous and 
infinitesimal (super)gauge symmetry transformations 
($\delta_{sg}, \delta_g$) (see, e.g. [6]):
\begin{eqnarray}
&&\delta_{sg} x_\mu = \kappa \;\psi_\mu, \qquad \delta_{sg} p_\mu = 0, \qquad \delta_{sg} \psi_\mu 
= i\; \kappa \; p_\mu, \qquad \delta_{sg} \chi = i\;\dot \kappa,
\nonumber\\ && \delta_{sg} e = 2 \;\kappa \; \chi, \qquad \delta_{sg} \psi_5 = i\;\kappa\;m, \qquad
\delta_g x_\mu = \xi \; p_\mu, \;\;\qquad \delta_g p_\mu = 0, \nonumber\\ &&
 \delta_g \psi_\mu = 0, \qquad
 \delta_g \chi = 0, \quad\qquad \delta_g e = \dot \xi,\quad\qquad \delta_g \psi_5 = 0, 
\end{eqnarray}
where $\kappa$ and $\xi$ are the (super)gauge infinitesimal parameters which are
fermionic (i.e. $\kappa^2 = 0, \kappa \psi_\mu + \psi_\mu \kappa = 0, \kappa \chi + \chi \kappa = 0$, etc.)
and bosonic (i.e. $ \xi^2 \neq 0, \xi \psi_\mu - \psi_\mu \xi = 0, \xi \chi - \chi \xi = 0$, etc.) in nature,
respectively. The generators (i.e. conserved charges) of the above {\it classical} continuous
transformations can be calculated in a straightforward manner. These are \footnote{ In the standard application of the Noether theorem, the (super)gauge
parameters  do {\it not} appear in the computation of the conserved  charge. However, we have kept these parameters
so that the connection between the generators (i.e. conserved charges) and the first-class constraints could be made clearly 
(in view of the seminal work done in Ref. [14]). }
\begin{eqnarray}
&&Q_{sg} = k\; (p\cdot\psi - m\;\psi_5 ),  \qquad \qquad 
Q_g = \frac{\xi}{2}\; \;(p^2 - m^2),
\end{eqnarray}
where the celebrated Noether's theorem has been exploited in its full blaze of glory. The above 
charges are the generators of  transformations (2) as can be checked  by using the standard formula:  
$\delta_\lambda \phi = \pm \; i\;[\phi, Q_\lambda]_\pm , (\lambda = sg, g)$ where the ($\pm$) signs, as the subscripts on the 
square bracket, correspond to the (anti)commutator for the generic variable $\phi$ being (fermionic)bosonic
in nature (and belonging to the Lagrangian (1)).

At this juncture, there are a few remarks in order. First, we note that the conserved charges in (3)
 are {\it unable} to generate the symmetry transformations
for the variables $e(\tau)$ and $\chi(\tau)$. Second, we point out that there 
is a reparametrization invariance in the theory because $L_f$ remains invariant under:  
$\delta_r x_\mu = \epsilon \; \dot x_\mu,\; \delta_r \psi_\mu = \epsilon \; \dot \psi_\mu,
\; \delta_r p_\mu = \epsilon \; \dot p_\mu, \; \delta_r \chi = \frac{d}{d\tau}(\epsilon \; 
\chi), \; \delta_r e = \frac{d}{d\tau}(\epsilon \; e)$,  
where $\epsilon$  is the infinitesimal transformation parameter  in: $\tau \rightarrow \tau^\prime 
= \tau - \epsilon (\tau)$. However, it can be shown that this transformation is equivalent 
to the gauge transformation in  specific limits [6]. Third, it can be shown that the
commutator of two supergauge transformations is equivalent to a reparametrization transformation
when the equations of motion, emerging from (1), are used (see, e.g. [3-5] for details).

In a very recent work [7], we have generalized the ``classical'' Lagrangian (1) and corresponding 
``classical'' symmetries to the
``quantum'' level within the framework of BRST formalism. The coupled (but equivalent) Lagrangians [that are consistent
generalizations  the Lagrangian (1)] are  (anti-)BRST invariant. These Lagrangians, 
in explicit form, are as follows  [7]   
\begin{eqnarray}
L_{\bar b} &=& L_f  - \bar b \;\dot e + \bar b\;(\bar b + 2\;\bar\beta\;\beta) + i\; \dot c\;(\dot{\bar c} 
+ 2\;\bar\beta\;\chi)   + 2\;i\;\beta \;\bar c\;\dot\chi \nonumber\\ &+& 2\;i\;\dot\beta \;\bar c\;\chi - 2
\;e\;(\gamma\;\chi - \beta\;\dot{\bar\beta}) + 2\;\beta\;\gamma\;\bar c  + \bar\beta^2\;\beta^2 + 
2\;\bar\beta\; c\;\gamma,
\end{eqnarray}
\begin{eqnarray}
L_b &=& L_f + b \;\dot e + b\;(b + 2\;\beta\;\bar\beta) - i\; \dot{\bar c}\;(\dot c 
+ 2\;\beta\;\chi) - 2\;i\;\bar\beta \; c\; \dot\chi
 \nonumber\\ &-& 2\;i\;\dot{\bar\beta} \; c\; \chi - 2\;e\;(\gamma\;\chi + \bar\beta\;\dot\beta) +
 2\;\beta\;\gamma\;\bar c + \bar\beta^2\;\beta^2 + 2\;\bar\beta\; c\;\gamma,
\end{eqnarray}
where $b$ and $\bar b$ are the Nakanishi-Lautrup type auxiliary variables, $(\bar \beta)\beta$ are the
bosonic (anti-)ghost variables in addition to the fermionic ($c^2 = \bar c^2 = 0,\; c\;\bar c 
+ \bar c\; c = 0$) (anti-)ghost variables $(\bar c)c$. We note that the (anti-)ghost variables $(\bar c)c$   are the generalizations of the bosonic  gauge parameter 
$\xi$ and the (anti-)ghost variables $(\bar \beta)\beta$ are needed for the supergauge parameter $\kappa$ 
[in transformations (2)] for the accurate derivation of the proper
(anti-)BRST invariant Lagrangians (4) and (5).
We also require a fermionic ($\gamma^2 = 0,  \gamma\chi + \chi\gamma  = 0, 
\gamma\psi_\mu + \psi_\mu \gamma = 0$, etc.) auxiliary 
variable $\gamma$ in the theory for the complete analysis of the above Lagrangians within the framework of BRST formalism.

The above {\it classical} infinitesimal 
transformations (2) can be generalized to the proper (i.e. off-shell nilpotent and absolutely 
anticommuting) (anti-)BRST symmetry transformations at the {\it quantum} level. 
The fermionic ``quantum'' 
(anti-)BRST symmetry transformations, corresponding to the {\it combined} ``classical'' ($\delta_g + \delta_{sg}$)  
transformations (2), are as follows (see, e.g. [7] for details)
\begin{eqnarray} 
&& s_{ab}\; x_\mu = {\bar c}\; p_\mu + \bar \beta \;\psi_\mu, \quad\qquad s_{ab}\; e 
= \dot {\bar c} + 2 \;\bar \beta\; \chi,  
\;\quad\qquad s_{ab} \;\psi_\mu = i \;\bar \beta\; p_\mu, \nonumber\\
&& s_{ab}\; \bar c = - i \;{\bar \beta}^2, \quad s_{ab}\; c = i\; \bar b, \quad s_{ab}\; \bar \beta = 0, 
\;\quad s_{ab} \; \beta = - i\; \gamma, \quad s_{ab}\; p_\mu = 0, \nonumber\\
&& s_{ab} \;\gamma = 0, \quad s_{ab}\; \bar b = 0, \quad s_{ab}\;\chi = i\; \dot {\bar \beta}, 
\quad s_{ab} \; b =  2\; i\; \bar \beta\; \gamma, \quad s_{ab}\psi_5 = i\;\bar\beta\; m,
\end{eqnarray}
\begin{eqnarray}
&&s_b\; x_\mu = c\;p_\mu + \beta \;\psi_\mu, \;\quad\qquad s_b\; e = \dot c + 2\;\beta\; \chi,  
\;\quad\qquad s_b\; \psi_\mu = i\;\beta\; p_\mu,\nonumber\\
&& s_b\;c = - i\; \beta^2, \;\;\quad s_b \;{\bar c} = i\; b, \;\;\quad s_b \;\beta = 0, 
\;\;\quad s_b \;\bar \beta = i \;\gamma, \;\;\quad s_b\; p_\mu = 0,\nonumber\\
&& s_b \;\gamma = 0, \quad s_b \;b = 0, \quad s_b \;\chi = i\; \dot \beta, 
\quad s_b\; \bar b = - 2\; i\; \beta\; \gamma, \quad s_b\;\psi_5 = i\;\beta\; m.
\end{eqnarray}
We lay emphasis  on the fact that, {\it only} for the combined ``classical'' ($\delta_g + \delta_{sg}$)
transformations, the off-shell nilpotent ``quantum'' BRST symmetries exist [6].
We note that the above transformations are off-shell nilpotent (i.e. $s_{(a)b}^2 = 0$) 
and their absolute anticommutativity ($s_b\;s_{ab} + s_{ab}\;s_b = 0$) is guaranteed  only when 
the following Curci-Ferrari (CF) type restriction, emerging from the superfield formalism
[7], namely;
\begin{eqnarray}  
b + \bar b + 2\;\beta\;\bar\beta = 0,
\end{eqnarray}
is satisfied. For instance, it can be checked explicitly  that $\{s_b, s_{ab}\}x_\mu = 0$ and $\{s_b, s_{ab}\}e = 0$ 
are true if and only if CF-type restriction (8) is  obeyed. Furthermore, this restriction
is found to be (anti-)BRST invariant 
(i.e. $s_{(a)b} [b + \bar b + 2\;\beta\;\bar\beta] = 0$) and it can  (besides superfield formalism 
[7]) be also obtained 
from the equations of motion: $b = - {\dot e}/{2} - \beta\bar\beta, \;\bar b =  
{\dot e}/{2} - \beta\bar\beta$ that emerge from the coupled (but equivalent) 
Lagrangians (4) and (5).

\section{Constraint analysis: A brief sketch}

As pointed out earlier, the starting Lagrangian (1)  is endowed with the first-class constraints: 
$\Pi_e \approx 0,\; \Pi_\chi \approx 0, \; (p^2 - m^2) \approx 0, \; (p \cdot\psi - m \psi_5 ) \approx 0$ 
where we have used the notation $p\cdot\psi = p_\mu \; \psi^\mu$ and the standard notation of ``$\approx$" 
as weakly zero in the language of Dirac's prescription for the constraint analysis. These constraints
are responsible for the (super)gauge transformations (2). However, as pointed out earlier, the 
conserved Noether charges (3) are {\it unable} to generate the (super)gauge transformations for 
$\chi$ and $e$ (which is a drawback in the Noether theorem for continuous symmetries for our present SUSY system).

One can write the generators of transformations (2), 
in the language of the first-class
constraints\footnote{It will be noted that the derivation of the expression for generator in Ref. [14]
is {\it not} for a SUSY system. For instance, the origin of the factor ($2 \kappa \chi$), associated with the primary 
constraints $\Pi_e$ in $G^{(sg)}$, is not discussed with any theoretical backing.}, as: (see, e.g. [14] for the detail discussions)  
\begin{eqnarray}
 G^{(sg)} =  \Pi_e\; (2 \;\kappa\; \chi) - i\;\Pi_\chi\; \dot \kappa +  {\kappa} \;(p \cdot \psi - m \; \psi_5),
\qquad G^{(g)} = \Pi_e \; \dot \xi + \frac{\xi}{2} (p^2 - m^2).
\end{eqnarray}   
It can be readily checked that, for the generic  variables $\phi$, we have the following:
\begin{eqnarray}
\delta_\lambda \phi = \pm \;i\; [\phi, G^{(\lambda)}]_{(\pm)} , \qquad\qquad \lambda = g, \;sg,
\end{eqnarray}    
where ($\pm$) signs, as the subscripts  on the square bracket, correspond to (anti)commutator for $\phi$ being 
(fermionic) bosonic in nature. The ($\pm$) signs, in front of the square bracket, are chosen judiciously 
(see, e.g. [15]). In the explicit computations of (10), the following basic canonical 
(anti)commutators have to be exploited:
\begin{eqnarray}
&& [e, \Pi_e] = i,\quad  \{\psi_5, \psi_5\} = 1,\quad  \{\chi, \Pi_\chi\} = i, \nonumber\\ 
&&[x_\mu, p_\nu] = i\;\eta_{\mu\nu}, \;\;\qquad \{\psi_\mu, \psi_\nu\} = -\;\eta_{\mu\nu}, 
\end{eqnarray} 
which emerge from the starting Lagrangian (1) (for $\hbar = 1$). The rest of the (anti)commutators are zero
for all the variables of our present SUSY system.

With the help of the beautiful relations in (9) and (10), we are able to be consistent with the Dirac's prescription 
for the quantization of system with constraints. For instance, at this stage, we are able to prove the time-evolution 
invariance of the first-class constraints of the theory.
In other words, the conditions $G^{(\lambda)}|phys> = 0$,  (where $|phys>$  are the physical states  of the theory), 
is able to lead to the following [cf. (9)].
\begin{eqnarray}
&& \dot \Pi_e \;|phys> = 0  \; \Rightarrow \; - \frac{1}{2}\;(p^2 -m^2)\; |phys> = 0,
 \nonumber\\ && \dot \Pi_\chi \;|phys> = 0  \; \Rightarrow   
\;i(p\cdot\psi -  m \psi_5 )\; |phys> = 0, \nonumber\\
&&\Pi_e \;|phys> = 0,\qquad  \qquad \Pi_\chi \;|phys> = 0, 
\end{eqnarray}         
in a clear and  consistent manner {\it at the same time}. This can be shown by using the equations of motion,
derived from Lagrangian (1), which imply the following
\begin{eqnarray}
&& \frac{d}{d \tau}  \Bigl(\frac{\partial L_f}{\partial \dot \chi} \Bigr) 
= \frac{\partial L_f} {\partial \chi} \Rightarrow 
\dot \Pi_\chi = \frac{\partial L_f} {\partial e} = i (p \cdot \psi  - m \psi_5), \nonumber\\
&& \frac{d}{d \tau}  \Bigl (\frac{\partial L_f}{\partial \dot e} \Bigr ) 
= \frac{\partial L_f} {\partial e} \Rightarrow 
\dot \Pi_e = \frac{\partial L_f} {\partial e} = -\frac{1}{2}  (p^2 - m^2).
\end{eqnarray}
Ultimately, we note that the generators $G^{(\lambda)} (\lambda = g, sg)$ 
(which are written in an ad-hoc fashion) are able to provide a consistent set 
of conditions on the physical states which are in total agreement with Dirac's prescription
for the quantization of system with constraints. However, a question
still remains to be answered. We have to provide a theoretical basis for the derivation of generators (9) {\it together} by
exploiting the basic principles of theoretical physics. This is what precisely we try to do in our next section by exploiting 
the BRST formalism.

\section{ Conserved BRST charge: As a generator}

We can address the above question within the framework of BRST formalism. To elaborate it, 
we note that the Lagrangians (4) and (5) transform as follows:
\begin{eqnarray}
 s_b\; L_b &=& \frac{d}{d\tau}\;\Bigl[\;\frac{1}{2}\;c\; (p^2 + m^2) + \frac{1}{2}\;\beta
\;(p \cdot\psi + m\;\psi_5 )  + \;b\;(\dot c + 2\; \beta\; \chi) 
\nonumber\\  &+&  2\; \gamma\;c\;\chi - 2\;\bar\beta\;\beta^2\;\chi 
- 2\;\bar\beta\;\dot\beta\;c \Bigr], \nonumber\\
  s_b\; L_{\bar b} &=& \frac{d}{d\tau}\;\Bigl[\;\frac{1}{2}\; c\; (p^2 + m^2) + \frac{1}{2}\; \beta 
(\;p \cdot\psi + m\;\psi_5)  + \;2\;i\;e\;\beta\;\gamma \nonumber\\ &-& 2\; b\; \beta\;\chi 
 -\bar b\;(\dot c + 2\;\beta\; \chi)
 + 2\;\beta\;\dot\beta\; \bar c \;\Bigr] \nonumber\\ &
+& (\dot c + 2\;\beta\;\chi)\;\frac{d}{d\tau}\;\Bigl[b + \bar b 
+ 2\;\beta\;\bar\beta \Bigr]  
 - (2\;i\;\beta\;\gamma)\; (b + \bar b + 2\;\beta\;\bar\beta),
\end{eqnarray}
under the BRST symmetry transformations (7). Thus, it is crystal clear that $L_b$ 
respects a {\it perfect} BRST symmetry but, under the very same BRST transformations,
$L_{\bar b}$ changes  to a total derivative plus terms that are zero on the hyper super world-line 
where the CF-type restriction (8) is valid. In other words, second and third terms of
(14) vanish due to the validity of (8) in the infinitesimal variation (i.e. $s_b L_{\bar b}$) of $L_{\bar b}$.

Exploiting the standard techniques of Noether's theorem in the context of action principle,
it can be seen that the following BRST charge:
\begin{eqnarray}
Q_b = \frac{1}{2}\; c\;(p^2 - m^2) + \beta\; (p\cdot\psi - m\;\psi_5 ) + b\; (\dot c +
2\;\beta\;\chi) + \beta^2 \;(\dot{\bar c}  + 2\;\bar\beta \;\chi),   
\end{eqnarray}
is conserved (i.e. ${\dot Q}_b = 0$) when we use the following equations of motion
\begin{eqnarray}
&&{\dot \psi}_\mu = \chi\;p_\mu,\qquad\qquad {\dot p}_\mu = 0,\qquad\qquad {\dot x}_\mu 
= e\;p_\mu - i\;\chi\; \psi_\mu, \qquad\qquad {\dot\psi}_5 = \chi\;m, \nonumber\\ 
&& \dot b + 2\;(\gamma\;\chi + \bar\beta\;\dot\beta) + \frac{1}{2} \;(p^2 - m^2) = 0, \qquad b\;\beta 
+ i\; \dot c\; \chi - e\;\dot\beta + \bar\beta\;\beta^2 + c\; \gamma = 0,\nonumber\\
&& \ddot c + 2\; \dot\beta\;\chi + 2\;\beta\;\dot\chi + 2\; i\;\beta\;\gamma = 0,  \;\;\;\quad b\;\bar\beta 
+ \dot e\; \bar\beta + e\; \dot{\bar\beta}  + \beta\;{\bar\beta}^2 
+ \gamma\; \bar c + i\; \chi \;\dot{\bar c} = 0,
\nonumber\\ 
&&\ddot{\bar c} + 2\; \dot{\bar\beta}\;\chi + 2\;\bar\beta\;\dot\chi + 2\; i\;\bar\beta\;\gamma = 0,
\qquad\qquad\qquad e\;\chi + \bar\beta \;c -\beta \; \bar c = 0, \nonumber\\ && (p \cdot\psi -  m \;\psi_5) 
+ 2\;\beta \;\dot{\bar c} - 2\;\bar\beta\; \dot c - 2\;i\;e\;\gamma = 0, 
\;\;\qquad \frac {\dot e}{2}  =  -\;(b + \beta\;\bar\beta), 
\end{eqnarray}
that emerge out from the Lagrangian $L_b$. We note that the BRST charge $Q_b$ is superior to 
the generator $G^{(\lambda)}$ [cf. (9)] because (i) it generates all the transformations of equation
 (7), (ii) it produces all the constraints (and 
their time-evolution invariance) through the physicality criteria where we demand that the true physical states are those that are annihilated by the BRST charge $Q_b$, and
(iii) there is a fundamental principle (i.e. Noether's theorem)  involved in its precise derivation.

To elaborate  on the above statements, it can be explicitly  checked that the analogue of equation (10), with
the BRST charge (15), produces all the transformations (7) for the {\it dynamical} variables. The transformations
for the {\it auxiliary} variables are obtained by the requirements of nilpotency and anticommutativity with 
anti-BRST symmetries (which we discuss in the next section). Finally, the physicality requirement with the 
conserved and nilpotent BRST charge $Q_b$, namely;
\begin{eqnarray}
Q_b\;|phys> = 0,
\end{eqnarray}
produces all the appropriate 
constraints (and their time-evolution invariance) on the physical states. Thus, we note that $Q_b$ is superior to the generators $G^{(\lambda)}$ [written in equation (9)], as we derive $Q_b$ by using  the Noether's theorem
which is a fundamental principle. We discuss the constraint analysis, in more detail, in the 
next section. We close this section with the remark that (17) does not lead to $\Pi_{\chi} |phys> = 0$. In this sense, the generators $G^{(\lambda)}$
are also, in some sense, superior to $Q_b$.

\section{Anti-BRST charge: Physicality criteria}

We focus now on the Lagrangians $L_{\bar b}$ and $L_b$ and discuss their anti-BRST symmetry 
invariance under the transformations (6). In fact, it can be readily 
checked [7] that we have the following transformations for the Lagrangian $L_b$ and $L_{\bar b}$, namely; 
\begin{eqnarray}
&& s_{ab}\; L_{\bar b} = \frac{d}{d\tau}\;\Bigl[\;\frac{1}{2}\;\bar c\; (p^2 + m^2)
+ \frac{1}{2}\; \bar\beta \;(\;p \cdot\psi + m\;\psi_5 )  
- \bar b\;(\dot {\bar c} + 2\; \bar\beta\; \chi) \nonumber\\ 
&& \quad\qquad +\; 2\;\gamma \;\bar c\;\chi + 2\;\beta\; \bar\beta^2\; \chi 
+ 2\;\beta\; \dot{\bar\beta}\;\bar c  \Bigr], \nonumber\\
&& s_{ab}\; L_ b = \frac{d}{d\tau}\;\Bigl[\;\frac{1}{2}\;\bar c\; (p^2 + m^2)
+ \frac{1}{2}\; \bar\beta (\;p \cdot\psi + m\;\psi_5)
+ 2\;i\;e\;\bar\beta\;\gamma  + 2\; \bar b\; \bar\beta\;\chi \nonumber\\ &&\quad\qquad +\;\; b\;(\dot {\bar c} + 2\; \bar\beta\; \chi) 
- 2\;\bar\beta \; \dot{\bar\beta} \; c \Bigr] + (2\;i\;\bar\beta\;\gamma)\;(b + \bar b + 2\;\beta\;\bar\beta)
\nonumber\\ && \quad\qquad - \;(\dot{\bar c} + 2\;\bar\beta\;\chi)\;\frac{d}{d\tau}\;\Bigl[b
 + \bar b + 2\;\beta\;\bar\beta \Bigr].
\end{eqnarray}
Thus, we note that $L_{\bar b}$ has a {\it perfect} symmetry under the nilpotent transformations $s_{ab}$ because
it transforms  to a total time derivative. On the contrary, under the very same symmetry transformations
 $s_{ab}$, the Lagrangian $L_{b}$ transforms to a total time derivative plus terms that 
are zero on the constrained hyper super world-line 
where the CF-type restriction $b + \bar b + 2\;\beta\bar\beta = 0$ is satisfied.
In other words, second and third terms, in
the variation $s_{ab} L_b$, would vanish due to the validity of (anti-)BRST invariant  CF-type restriction (8).

According to Noether's theorem, the invariance of action $S = \int d\tau L_{\bar b}$, under continuous 
and nilpotent symmetry transformations $s_{ab}$, leads to the derivation of  nilpotent 
(i.e. $Q_{ab}^2 = 0$) and conserved (i.e. $\dot Q_{ab} = 0$) charge $Q_{ab}$, in an accurate fashion,
as follows 
\begin{eqnarray}
Q_{ab} &=& \frac{1}{2}\; \bar c\;(p^2 - m^2) + \bar\beta\; (p\cdot\psi - m\;\psi_5 )  
- \bar b\; (\dot {\bar c} + 2\;\bar\beta\;\chi) - \bar\beta^2 \;(\dot c + 2\;\beta \;\chi).   
\end{eqnarray}
The conservation law (i.e. ${\dot Q}_{ab} = 0$) of this charge $Q_{ab}$ 
can be proven by exploiting the following Euler-Lagrange equations of motion 
\begin{eqnarray}
&&{\dot x}_\mu = e\;p_\mu - i\;\chi\; \psi_\mu,\qquad {\dot \psi}_\mu = \chi \;p_\mu, \qquad \dot {\bar b} = 2\;(\gamma\;\chi - \beta\;\dot{\bar\beta}) + \frac{1}{2} \;(p^2 - m^2), \nonumber\\
&& \bar b\;\bar\beta - i\; \dot{\bar c}\; \chi + e\;\dot{\bar\beta} + \beta\;\bar\beta^2 
+ \gamma\;\bar c  = 0, \;\;\qquad e\;\chi - \beta \;\bar c +\bar\beta \; c = 0,\;\qquad {\dot p}_\mu = 0,
 \nonumber\\ && \ddot c + 2\; \dot\beta\;\chi + 2\;\beta\;\dot\chi 
+ 2\; i\;\beta\;\gamma = 0,\qquad \qquad
  \ddot{\bar c} + 2\; \dot{\bar\beta}\;\chi + 2\;\bar\beta\;\dot\chi 
+ 2\; i\;\bar\beta\;\gamma = 0,\nonumber\\ &&{\dot\psi}_5 = \chi \; m, \qquad\qquad 
(p \cdot\psi - m\;\psi_5) + 2\;\beta \;\dot{\bar c} - 2\;\bar\beta\; \dot c
 - 2\;i\;e\;\gamma = 0, \nonumber\\ &&  \dot e\; \beta + e\; \dot\beta 
 - \bar b\;\beta - \bar\beta\;\beta^2 + \gamma\; c + i\; \chi \;\dot c = 0,\quad\qquad
\frac{\dot e}{2} = (\bar b + \;\beta\;\bar\beta), 
\end{eqnarray}
that emerge out from $L_{\bar b}$.

Exploiting the generator equation (10) appropriately, it is straightforward to check that the conserved charge 
$Q_{ab}$ is the generator of the transformations $s_{ab}$ [cf. (6)]. It is worthwhile to point out  that $Q_{ab}$ generates transformations 
$s_{ab}$ {\it only} for the dynamical variables of the theory. Such transformations for the auxiliary variables (e.g. $b, \bar b, \gamma$, etc.)
are obtained by the requirements of nilpotency, absolute anticommutativity and (anti-)BRST invariance of the CF-type restriction (8). Furthermore, 
the physicality criteria (17) with the anti-BRST charge  $Q_{ab}$ leads to the annihilation of {\it all} the physical state $|phys>$ 
by the operator forms of the first-class constraints $\Pi_e \approx 0, \;(p^2 - m^2) \approx 0,
\;(p\cdot\psi - m \psi_5 )\approx 0$. This observation is consitent with the requirements of Dirac's prescription for quantization of  
physical systems with any arbitrary kinds of constraints.

To corroborate  the above statements, it is essential to use the explicit expressions for $Q_{ab}$ and equations of motion (20) that have 
been derived from $L_{\bar b}$. In fact, the physicality criteria $Q_{ab} |phys> = 0$ implies the following
\begin{eqnarray}
&&(p^2 - m^2)\;|phys> = 0 \quad\quad\Rightarrow   \dot \Pi_e \;|phys> = 0, \nonumber\\&&
(p\cdot \psi -\psi_5\; m)\;|phys> = 0 \Rightarrow   {\dot \Pi}_\chi \;|phys> = 0, \nonumber\\
&&\bar b\;  |phys> = 0 \;\;\quad\qquad\qquad\Rightarrow  \Pi_e \;|phys> = 0,
\end{eqnarray}
 We note that $\Pi_\chi |phys> = 0$ is {\it not} produced by the physicality criteria $Q_{(a)b} |phys> = 0$
with the conserved and nilpotent (anti-)BRST charges $Q_{(a)b}$.
We elaborate on this {\it new} feature as well as compare and contrast the importance of  $G^{(\lambda)}$ 
and $Q_{(a)b}$  in our ``conclusions" section (see, Sec. 7 below). Thus, we clearly note that the 
(anti-)BRST charges $Q_{(a)b}$ are  theoretically more appealing  than the straightforward  
calculations of generators $G^{(\lambda)}$ ($\lambda = g, sg$) [cf. (9)] as there is a consistent theoretical basis for their derivations.

\section{Ghost symmetry: Conserved ghost charge and BRST algebra}

It can be readily checked that if the (anti-)ghost variables $(\bar c)c$ and $(\bar\beta)\beta$
undergo the following scale transformations [with $\Omega$ as a  global  (i.e. spacetime independent) 
scale parameter]:
\begin{eqnarray}
&& c \rightarrow  e^{+ \Omega} \;c, \;\qquad \bar c \rightarrow  e^{-\Omega} \; \bar c, \;\qquad
\beta \rightarrow  e^{+\Omega} \;\beta, \;\qquad \bar\beta \rightarrow  e^{-\Omega} \bar\beta,\nonumber\\
&& (x_\mu, \psi_\mu, e, \chi, b, \bar b, \gamma)\rightarrow e^0\;  
(x_\mu, \psi_\mu, e,  \chi, b, \bar b, \gamma),
\end{eqnarray}
the (anti-)BRST invariant Lagrangians (4) and (5) remain invariant. The infinitesimal version
of the ghost transformations (22) are as follows
\begin{eqnarray}
&& s_{gh} c = c,\;\qquad s_{gh} \bar c = - \bar c,\;\;\qquad s_{gh} \beta = \beta,\nonumber\\ &&
s_{gh} \bar\beta = - \bar\beta, \quad s_{gh} (x_\mu, \psi_\mu, e, \chi,  b, \bar b, \gamma) = 0,
\end{eqnarray}
which are derived after setting $\Omega = 1$ for the sake of brevity.
The numbers in the exponentials denote the ghost numbers of the variables. We further note that the ghost
numbers for ($x_\mu,\;\psi_\mu,\;e,\; \chi, \;b,\; \bar b, \gamma$) are zero. As  a result, these variables 
do {\it not} transform under $s_{gh}$.

At this stage, a few comments are in order as far as the symmetry operators $\delta_g, \delta_{sg}, s_b, s_{ab}$
and $s_{gh}$ are concerned. First, it can be explicitly checked that $\delta_g$ and $\delta_{sg}$
are  independent of each-other because $[\delta_g, \delta_{sg}]  = 0$ when they operate on any arbitrary  variable of the Lagrangian (1). Second, the symmetry operators $s_b, s_{ab}$ and $s_{gh}$ obey the following algebra,  namely; 
 \begin{eqnarray}
&& s_b^2 = 0, \quad\qquad s_{ab}^2 = 0,\quad\qquad  \{s_b, s_{ab}\} = 0, \nonumber\\ && [s_{gh}, s_b] = + s_b, 
\;\;\qquad\qquad [s_{gh}, s_{ab}]  =  - s_{ab}.
\end{eqnarray}
We would like to point out that, for the proof of absolute  anticommutativity
property of (anti-)BRST symmetry transformations (i.e. $s_b \;s_{ab} + s_{ab}\;s_b = \{s_b, s_{ab}\} = 0$),
one has to exploit the (anti-)BRST invariant CF-type restriction (8) in an explicit fashion.

According to Noether's theorem, the invariance of the Lagrangians (4) and (5) under the infinitesimal version of ghost-scale transformations (22) leads to the derivation of the following conserved ghost charge ($Q_{gh}$): 
\begin{eqnarray}
Q_{gh} = 2\;i\; \beta\; \bar c\; \chi + 2\;i\;\bar\beta\;c\;\chi - 2\;e\;\beta\;\bar\beta.
\end{eqnarray}
The conservation law (i.e. $\dot Q_{gh} = 0$) of the above charge can be proven by exploiting the appropriate equations of motion
derived from the Lagrangians (4) and (5). The conserved charges $Q_g$ and $Q_{sg}$ [cf. (3)]  commute with 
each-other (i.e. $[Q_{g}, Q_{sg}] = 0$). The rest of the conserved charges $Q_b, Q_{ab}$ and $Q_{gh}$ 
obey the following standard BRST algebra   
\begin{eqnarray}
&& Q_b^2 = 0, \qquad Q_{ab}^2 = 0,\qquad  \{Q_b, Q_{ab}\} = 0, \nonumber\\ && i\;[Q_{gh}, Q_b] =  + Q_b,
\quad i\;[Q_{gh}, Q_{ab}]  =  - Q_{ab},
\end{eqnarray}
which establishes the nilpotency of (anti-)BRST charges $Q_{(a)b}$ and the fact that the 
ghost numbers for the nilpotent (anti-)BRST charges $Q_{(a)b}$ are ($\mp 1$), respectively.

The above proper BRST algebra (26) can be obtained by using the canonical (anti)commutators, derived from the 
Lagrangians (4) and (5). There is  a simpler  way to derive these relations
where the generator equation (10)
plays an important role. For instance, the following is true [if we use (16) and (20)]:
\begin{eqnarray}
&&s_b\;Q_b \;= -\; i\;\{ Q_b, Q_b \} =  0, \qquad
s_{ab}\;Q_{ab} = -\; i\;\{ Q_{ab}, Q_{ab} \} =  0, \nonumber\\
&&s_b Q_{ab} = -\; i\;\{ Q_{ab}, Q_b \} =  0, \qquad
s_{ab} Q_b = -\;i\; \{ Q_b, Q_{ab} \} =  0,\nonumber\\
&&s_{gh} Q_b = +\; i\;[Q_{gh}, Q_b ] =  Q_b, \qquad 
s_{gh} Q_{ab} = +\; i\; [Q_{gh}, Q_{ab} ] = - Q_{ab},   
\end{eqnarray}
where the l.h.s. uses the transformations (6), (7) and (23) and expressions (15) and (19) in 
their explicit forms.
The above relations capture the BRST algebra (26). We would like to lay emphasis on the
fact that, in the proof of anticommutativity $s_b Q_{ab} \equiv s_{ab} Q_b = \{Q_b, Q_{ab} \} = 0$, we have to use (8).

\section{Conclusions}

In our previous paper [7] and in our present investigation, we have observed
many novel features associated with the model of a spinning relativistic particle
when it is considered within the framework of BRST formalism. For instance, the
existence of the coupled (but equivalent) Lagrangians, derivation of the proper
(i.e. off-shell nilpotent and absolutely anticommuting) (anti-)BRST symmetries, 
emergence of the (anti-)BRST invariant CF-type restriction, etc., are some 
of the completely new results for this model which are elucidated, for the first-time,
in our present and earlier [7] works.

We observe, for the first-time, that the physicality criteria with the conserved
and nilpotent (anti-)BRST charges do {\it not} produce the annihilation of the
physical states by one of the primary constraints of the theory. In other words,
we find that $Q_{(a)b} |phys> = 0$ does not imply $\Pi_{\chi} |phys> = 0$, 
in our present model, due to the  fact that $\Pi_{\chi} =  2 i \bar \beta c$
(and/or $\Pi_{\chi} = - 2 i \beta \bar c$) is accurately  
expressed {\it only} in terms of
the (anti-)ghost variables (cf. $L_b$ and $L_{\bar b}$). 
As a consequence, it is but natural
to find that the primary constraint ($\Pi_{\chi} |phys> = 0$) does {\it not}
appear\footnote{The other way of saying this fact is that
$\Pi_{\chi}$ is nothing but the (anti-)ghost variables  which are not the
physical objects of our present theory. As a result, the constraint condition $\Pi_\chi |phys> = 0$ does {\it not} ensue from the above requirements of 
the physicality criteria ($Q_{(a)b}|phys> = 0$).}   from the physicality criteria $Q_{(a)b} |phys> = 0$. 
The other way of saying this fact is that
$\Pi_{\chi}$ is nothing but the (anti-)ghost variables  which are not the
physical objects of our present theory. As a result, $\Pi_\chi |phys> = 0$ does {\it not} ensue from the above requirements of 
the physicality criteria (i.e. $Q_{(a)b}|phys> = 0$).

The deeper theoretical reason  behind the above riddle is the fact that the kinetic term for $\chi$ variable 
does  {\it not} exist (i.e. $\dot\chi^2 = 0$) unlike the case of  the bosonic {\it gauge} variable where it does
[i.e. $(- \dot e^2 /2) = b \;\dot e + b^2 /2$]. In fact, this is the reason that, ultimately, 
in the expression  for $Q_b$, we have a term $b(\dot c + 2 \beta \chi)$ [cf. equation (15)]
and $Q_b |phys> = 0$ implies that $b|phys> = 0$  which is equivalent to the statement  
$\Pi_e |phys>  = 0$ (as $b = \Pi_e$). No such thing happens for $\Pi_\chi$ 
as there is no analogue of ``$b$'' in the expression for $Q_b$ as the momentum for $\chi$. 
Whether this features is a decisive property of a SUSY gauge theory, within  the framework of BRST formalism, we do not 
know at present. It is an open problem for us for the future investigations.

We observe that the conditions $G^{(\lambda)} |phys> = 0$ leads to the annihilation of the physical
states by the primary as well as secondary constraints {\it together} [cf. (12)]. However, the expression
for the $G^{(\lambda)}$ is somewhat ad-hoc. In  fact, the expression  for $G^{(\lambda)}$ ($\lambda = g, sg$) has been written because
we already know the symmetry transformations (2) and the first-class constraints
of the theory. It has not been derived from any basic principles. On the contrary, the physicality criteria with (anti-)BRST
charges ($Q_{(a)b} |phys> = 0$) do {\it not} yield $\Pi_\chi |phys> = 0$ but $Q_{(a)b}$ are derived from
the basic principles. Logically one can explain the {\it non-existence} of $\Pi_\chi |phys> = 0$, within
the framework of BRST formalism, because $\Pi_\chi =  2 i \bar \beta c$ (or $\Pi_\chi = - 2 i \beta \bar c$)
is expressed {\it only} in terms of the (anti-)ghost variables which are {\it non-physical}. Thus, we note
that $G^{(\lambda)}$ and $Q_{(a)b}$ have their own virtues and vices.

It would be very interesting endeavor for us to apply the BRST approach to find out the problems of quantization
associated with the pseudoclassical description of the massive spinning particle in {\it odd} and any arbitrary
dimension of spacetime [16-18]. Furthermore, we plan to apply our formalism to the issues of quantization
related with the field theoretic models of Chern-Simons theories with P-T invariance (see, e.g. [19,20]
and references therein). It would be very important for us to apply the superfield formalism [8-11] to
the description of above models so that we could find out the proper (anti-)BRST symmetries for
these theories.  This exercise will enable us to obtain the (anti-)BRST invariant Lagrangians (and
Lagrangian densities) of the above theories which, in turn, would lead to the discussions
about the quantization issues within the framework of BRST formalism.

It would be a very challenging problem to apply the key ideas of superfield
and BRST formalisms to other phenomenologically realistic SUSY models of gauge
theories so that some novel features could be explored
in physical four dimensions of spacetime. Further, we plan to address the problem of anomalies associated
with the spinning particles [21] within the framework of BRST formalism at the quantum level.  
It is gratifying to state, at this juncture, that we have already shown
the time-evolution invariance of the CF-type restriction within the framework of Hamiltonian
formalism in our recent publication [22]. We are currently deeply involved with 
{\it all} the above cited problems and
our results will be reported in our future publications [23]. 
\\

\noindent
{\bf Acknowledgments:}
Two of us (S. K. and D. S.) remain grateful to UGC, Govt. of India, New Delhi, for 
financial support under RGNF and RFSMS schemes. Fruitful comments by our esteemed
Referee are thankfully acknowledged, too.

\end{document}